\documentclass[9pt,twocolumn,twoside]{osajnl}

\journal{josab}

\setboolean{shortarticle}{false}

\usepackage{upgreek}
\usepackage{enumitem}

\usepackage{acronym}
\usepackage[caption=false]{subfig}

\acrodef{ADU}[ADU]{analog-digital unit}
\acrodef{AO}[AO]{adaptive optics}
\acrodef{ASM}[ASM]{adaptive secondary mirror}
\acrodef{CCD}{charged-coupled device}
\acrodef{CoM}{center of mass}
\acrodef{DARC}[DARC]{Durham Adaptive optics Real-time Controller}
\acrodef{DAC}[DAC]{digital-analog converter}
\acrodef{DIMM}{differential image motion monitor}
\acrodef{DM}[DM]{deformable mirror}
\acrodef{DFT}[DFT]{discrete Fourier transform}
\acrodef{ExAO}[ExAO]{extreme adaptive optics}
\acrodef{ELT}[ELT]{extremely large telescope}
\acrodef{FLAO}{First Light AO}
\acrodef{FMF}[FMF]{few-mode fiber}
\acrodef{FoV}{field of view}
\acrodef{FRD}[FRD]{focal ratio degradation}
\acrodef{FWHM}[FWHM]{full width at half maximum}
\acrodef{GPI}[GPI]{Gemini Planet imager}
\acrodef{IFU}[IFU]{integral field spectroscopy}
\acrodef{ISYS}[ISYS]{Institute for System Dynamics Stuttgart}
\acrodef{KIT}[KIT]{Karlsruhe Institute for Technology}
\acrodef{KOOL}[KOOL]{Koenigstuhl Observatory Opto-Mechatronics Laboratory}
\acrodef{LBT}[LBT]{Large Binocular Telescope}
\acrodef{LBTI}[LBTI]{Large Binocular Telescope Interferometer}
\acrodef{LBTI-AO}{\acl{LBTI} \acl{AO}}
\acrodef{LSW}[LSW]{Landessternwarte}
\acrodef{MCF}[MCF]{multi-core fiber}
\acrodef{MFD}[MFD]{mode-field diameter}
\acrodef{MLA}[MLA]{micro-lens array}
\acrodef{MLR}[MLR]{micro-lens ring}
\acrodef{MLR-TT}[MLR-TT]{micro-lens ring tip-tilt}
\acrodef{MM}[MM]{multi-mode}
\acrodef{MMF}[MMF]{multi-mode fiber}
\acrodef{MPIA}[MPIA]{Max Plank Institute for Astronomy}
\acrodef{NA}[NA]{numerical aperture}
\acrodef{NCP}[NCP]{non-common path}
\acrodef{NCPA}[NCPA]{non-common path aberration}
\acrodef{NAIR}[NAIR]{Novel Astronomical Instrumentation based on photonic light Reformating}
\acrodef{NIR}[NIR]{near-infrared}
\acrodef{PL}[PL]{photonic lantern}
\acrodef{POP}[POP]{\textit{Physical Optics Propagation}}
\acrodef{PSF}[PSF]{point-spread function}
\acrodef{PSD}[PSD]{power spectral density}
\acrodef{RMS}[RMS]{root mean square}
\acrodef{RV}[RV]{radial velocity}
\acrodef{SM}[SM]{single-mode}
\acrodef{SMF}[SMF]{single-mode fiber}
\acrodef{SNR}[SNR]{signal-to-noise ratio}
\acrodef{SR}[SR]{Strehl ratio}
\acrodef{VLT}[VLT]{Very Large Telescope}
\acrodef{VLTI}[VLTI]{Very Large Telescope Interferometer}
\acrodef{WFS}[WFS]{wavefront sensor}

\newacroindefinite{MLR}{an}{a}
\newacroindefinite{MMF}{an}{a}
\newacroindefinite{SMF}{an}{a}
\newacroindefinite{MFD}{an}{a}
\newacroindefinite{RMS}{an}{a}
\newacroindefinite{SR}{an}{a}
\newacroindefinite{MLA}{an}{a}
\newacroindefinite{DAC}{a}{a}

\newcommand{\mum}{\upmu \mathrm{m}}
\newcommand{\lambdaD}{\uplambda / \mathrm{D}}

\newcommand{\old}[1]{}

\title{On-sky results for the novel integrated micro-lens ring tip-tilt sensor}

\author[1,*]{Philipp Hottinger}
\author[1,2]{Robert J. Harris}
\author[3]{Jonathan Crass}
\author[4,5,6]{Philipp-Immanuel Dietrich}
\author[4,5]{Matthias Blaicher}
\author[3]{Andrew Bechter}
\author[3]{Brian Sands}
\author[7]{Tim J. Morris}
\author[7]{Alastair G. Basden}
\author[7]{Nazim Ali Bharmal}
\author[1]{Jochen Heidt}
\author[1,2]{Theodoros Anagnos}
\author[8]{Philip L. Neureuther}
\author[8]{Martin Gl{\"u}ck}
\author[9]{Jennifer Power}
\author[2]{J{\"o}rg-Uwe Pott}
\author[4,5,6]{Christian Koos}
\author[8]{Oliver Sawodny}
\author[1]{Andreas Quirrenbach}

\affil[1]{Landessternwarte, Zentrum f\"ur Astronomie der Universit\"at Heidelberg, K\"onigstuhl 12, 69117 Heidelberg, Germany}
\affil[2]{Max-Planck-Institute for Astronomy, K\"onigstuhl 17, 69117, Heidelberg, Germany}
\affil[3]{Department of Physics, University of Notre Dame, 225 Nieuwland Science Hall, Notre Dame, IN~46556, USA}
\affil[4]{Institute of Microstructure Technology~(IMT), Karlsruhe Institute of Technology (KIT), Hermann-von-Helmholtz-Platz~1, 76344~Eggenstein-Leopoldshafen, Germany}
\affil[5]{Institute of Photonics and Quantum Electronics (IPQ), Karlsruhe Institute of Technology (KIT), Engesserstr. 5, 76131~Karlsruhe, Germany}
\affil[6]{Vanguard Photonics GmbH, Hermann-von-Helmholtz-Platz 1, 76344~Eggenstein-Leopoldshafen, Germany}
\affil[7]{Department of Physics, Durham University, South Road, Durham, DH1 3LE, UK}
\affil[8]{Institute for System Dynamics, University of Stuttgart, Waldburgstr. 19, 70563 Stuttgart, Germany}
\affil[9]{Large Binocular Telescope Observatory, 933 N. Cherry Ave, Tucson, AZ 85721-0009, U.S.A.}

\affil[*]{phottinger@lsw.uni-heidelberg.de}

\affil[ ]{\textcopyright~2021 Optical Society of America]. One print or electronic copy may be made for personal use only. Systematic reproduction and distribution, duplication of any material in this paper for a fee or for commercial purposes, or modifications of the content of this paper are prohibited.}

\begin{abstract}

We present the first on-sky results of the \acl{MLR-TT}
(MLR-TT) sensor. This sensor utilizes a 3D printed \acl{MLR}
feeding six \aclp{MMF} to sense misaligned light, allowing
centroid reconstruction. A tip-tilt mirror allows
the beam to be corrected, increasing the amount of light
coupled into a centrally positioned \acl{SM} (science) fiber.
The sensor was tested with the iLocater acquisition
camera at the \acl{LBT} in November 2019. The limit on the
maximum achieved \acl{RMS} reconstruction accuracy was found
to be  0.19$\,\lambdaD$ in both tip and tilt, of which
approximately 50\% of the power originates at frequencies below 10\,Hz.
We show the reconstruction accuracy is highly dependent
on the estimated \acl{SR} and simulations support the
assumption that residual \acl{AO} aberrations are the main
limit to the reconstruction accuracy.
We conclude that this sensor is ideally suited to remove post-adaptive optics non-common path tip tilt residuals.
We discuss the next steps for the concept development,
including optimizations of the lens and fiber, tuning of the
correction algorithm and selection of optimal science cases.
\end{abstract}

\acresetall 

\setboolean{displaycopyright}{true}

\begin{document}

\maketitle

\section{Introduction}
\label{sec:intro}

In recent decades, improvements in the performance of an increasing number of \ac{ExAO} systems has led to the ability to image near the diffraction-limit using 8\,m-class telescopes  \cite{Beuzit2008,Esposito2011,Macintosh2014,Jovanovic2015}.
These \ac{ExAO} systems focus on achieving the best performance over a small \ac{FoV} and regularly achieve \acp{SR} of 80\% in the \ac{NIR}.
One of the most prominent goals for these systems is the direct observation and characterization of exoplanets \cite{Angel1994}, for which high angular resolution and contrast are crucial.
The high level of correction provided by these \ac{ExAO} systems also makes it possible to efficiently couple light from the telescope directly into \acp{SMF} \cite{Bechter2016}.
\acp{SMF} have a core diameter of the order of ten microns, which can only transport the fundamental fiber mode. As this mode is the only spatial mode transported and has a near-Gaussian intensity profile, the corresponding output beam is very stable and easy to model. \acp{SMF} also act as a spatial filter and  couple very little sky background \cite{Crepp2016}. This makes them highly suitable for direct exoplanet spectroscopy \cite{haffert2020} and interferometry \cite{CoudeduForesto1994,LeBouquin2011,Gillessen2010,Cvetojevic2018,Martinod2018}.
When coupled to a high resolution spectrograph, \acp{SMF} also remove conventional modal noise, allowing an increase in the achievable \ac{RV} precision \cite{Cvetojevic2017}.
A number of \ac{SMF}-fed spectrographs are currently under development, including iLocater at \ac{LBT} \cite{Crepp2016,Crass2020}, SPHERE and CRIRES+ at the \ac{VLT} \cite{Vigan2018}, RHEA and IRD at SCExAO/Subaru \cite{Feger2014,Kotani2018} and KPIC at Keck \cite{Mawet2016}.

\begin{figure}[htbp]
  \centering
  \includegraphics[width=\columnwidth]{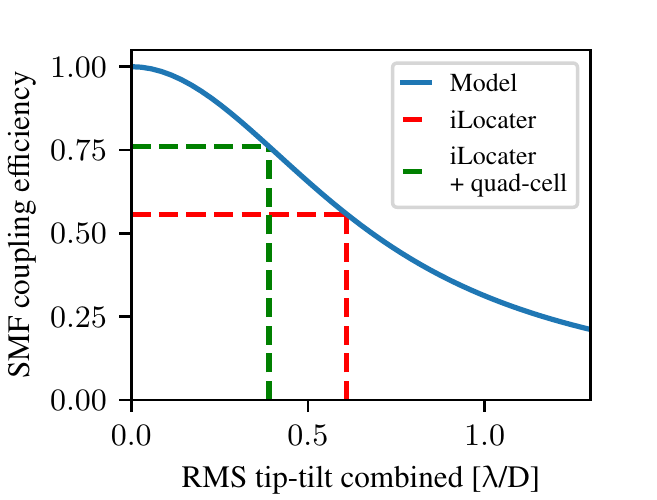}
  \caption{Numerically calculated theoretical normalized coupling efficiency assuming an optimally coupled diffraction-limited PSF with additional residual tip-tilt variation, plotted in units of $\lambdaD$.
  The measured RMS residuals at the iLocater focal plane are also indicated, without beam stabilization at 0.61\,$\lambdaD$ resulting in a theoretical reduction by 44\% (red line), and with additional stabilization with a quad-cell detector improving tip-tilt stability to 0.39\,$\lambdaD$, leading to a tip-tilt induced coupling loss of 24\% (green line) \cite{Crass2020}. }
  \label{fig:rms_coupling}
\end{figure}

As the size of the fiber is of the order of the diffraction limit ($\lambdaD$, where $\uplambda$ is the wavelength and D the diameter of the telescope), the alignment accuracy is highly dependent on the \ac{PSF} stability (see  Fig.~\ref{fig:rms_coupling} for an example of relative coupling efficiency as a function of residual tip-tilt position).
Any vibrations that occur throughout the telescope system and influence the position of the \ac{PSF} in the focal plane can have a large impact on performance. These variations can be caused by electrical and mechanical components such as fans and pumps, but can also be induced by wind, atmospheric distortions and dome seeing \cite{Milli2018}.
As these variations can have both large amplitude and high frequencies, the \ac{AO} system may not be able to compensate for them sufficiently and, if they can occur outside the path to the \ac{WFS}, they will not be sensed.
These variations can effect the performance significantly \cite{Lozi2018} and turn out to be a limiting factor when coupling into \acp{SMF}, with coupling efficiency being degraded by as much as a factor of two \cite{Bechter2019}.

Besides high-order \ac{AO} correction, efficient \ac{SMF}-coupling therefore requires a method to accurately sense and correct induced tip-tilt variations.
Traditionally, this is accomplished by detecting the \ac{PSF} at the focal plane either with a fast quad-cell photo detector \cite{Esposito1997} or camera, computing the centroid position, and feeding back a corresponding error signal to a fast tip-tilt correction mirror.
More advanced systems include feed-forward correction of mechanical vibration measurements with accelerometers \cite{Gluck2017} and the deployment of complex metrology systems utilizing concurrent alignment lasers \cite{Lippa2016}.
While most of these systems have been adopted at large telescopes, they all have a significant mechanical and optical footprint, throughput loss, tend to become complex in operation, and are vulnerable to \ac{NCP} effects as the tip-tilt correction is performed at a different optical surface than the \ac{SMF} face.

Different fiber based photonic sensor concepts are being investigated in the community to complement conventional \ac{AO} systems \cite{Corrigan:2016,norris2020all}. The concept presented in this work draws from Ref.~\citenum{Dietrich2017}, who developed a sensor with multiple \ac{SM} cores equipped with an \acl{MLA} to refract the beam at the focal plane for both science instrument and tip-tilt sensing.
Our modified concept features \acp{MMF} in conjunction with \iac{MLR} \cite{Hottinger2018} for sensing and is called the \acl{MLR-TT} (MLR-TT) sensor \cite{Hottinger2019}.
We present first on-sky results of this novel tip-tilt sensor with the iLocater acquisition camera at the \ac{LBT} \cite{Crass2020}.

In Section \ref{sec:methods}, we describe the sensor concept and the methods used to design, manufacture and employ it at the telescope along with outlining our simulation approach.
In Section \ref{sec:results} we present our on-sky results and supporting simulations and in Section \ref{sec:discussion} we discuss these results and future developments before presenting our conclusions in Section \ref{sec:conclusion}.

\section{Design and methods}
\label{sec:methods}

\begin{figure}[htbp]
  \centering\includegraphics[width=\columnwidth]{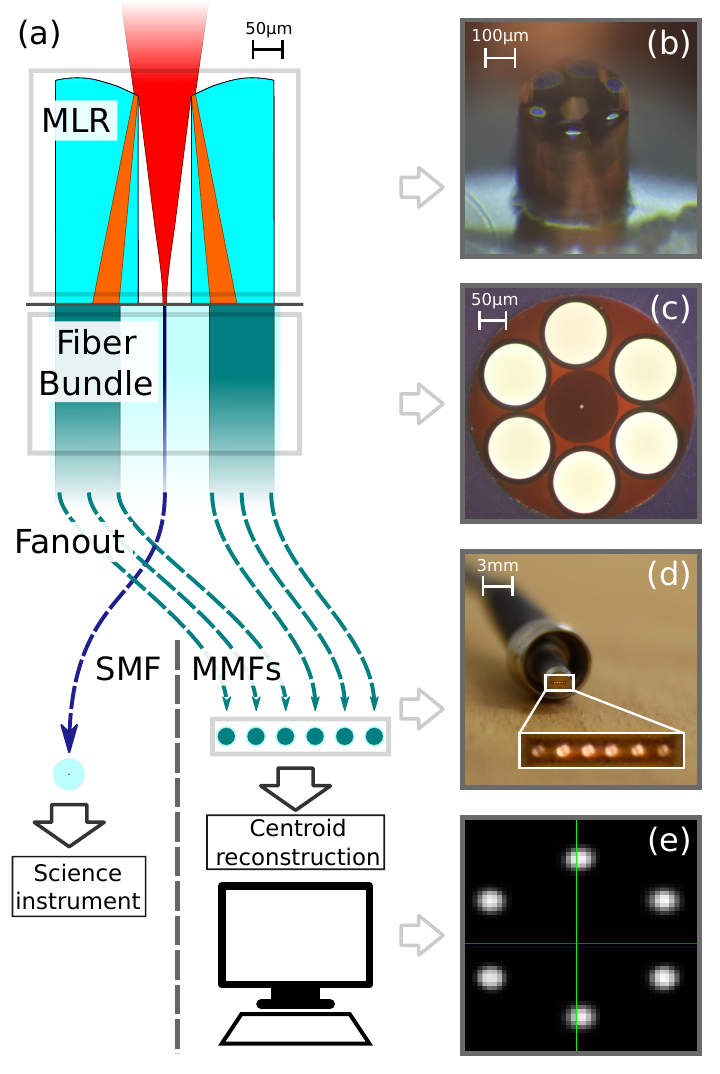}
  \caption{Overview of the micro-lens ring tip-tilt sensor (MLR-TT). (a) Schematics of the setup. The starlight (red) is coupled into the \ac{SMF} (dark blue), while some light at the edges of the beam is clipped and refracted (orange beam) by the \ac{MLR} (light blue) to be coupled into the sensing \acp{MMF} (dark green). The fibers are embedded in a fiber bundle that fans out into a single \ac{SMF} which then feeds the starlight into a science instrument and the six \acp{MMF} that are reformatted into a linear array mounted in an SMA connector. The sensing fibers are then re-imaged and the detected flux is used to reconstruct the centroid position of the telescope beam.
  (b)\,Microscope image of the \ac{MLR} on the fiber bundle face, (c)\,microscope image of back-illuminated fiber bundle, (d)\,sensing fiber output at the fiber connector, and (e)\,re-arranged detector signal for visual examination of the reconstruction algorithm with the green cross indicating the centroid position.}
  \label{fig:baremlr}
\end{figure}

The MLR-TT sensor concept is depicted in Fig.~\ref{fig:baremlr} as both a schematic cross-section of the optics (Fig.~\ref{fig:baremlr}, left-hand side) and as images of the manufactured components (Fig.~\ref{fig:baremlr}, right-hand side). The details are re-iterated here, with additional information, for clarity:

\begin{enumerate}[itemsep=0mm]
    \item The sensor consists of a fiber bundle containing six \acp{MMF} surrounding \iac{SMF}, located at the iLocater focal plane. On the fiber face, \iac{MLR} stands 380\,$\mum$ tall and 355\,$\mum$ wide with a central aperture of 86\,$\mum$.
    \item The central part of the beam is injected into the \ac{SMF}, while the outer edge is clipped and refracted by the \ac{MLR}. Depending on the alignment of the beam, the proportion of light clipped by the \ac{MLR} changes, which modifies the coupling into the individual \acp{MMF}.
    \item The \acp{MMF} are separated from the \ac{SMF}, re-arranged to form a linear array, re-imaged, and read out by a detector.
    \item The illumination pattern of the \acp{MMF} is processed to reconstruct the original \ac{PSF} centroid position, which can be fed back to a fast steering tip-tilt mirror for correction.
\end{enumerate}

\subsection{Fiber bundle design}
\label{sec:methods-fiber-methods}

The fiber bundle was manufactured commercially (Berlin Fibre GmbH) and holds the array of seven fibers terminated into an FC/PC connector which is then connected to the iLocater fiber feed mount.
The fibers are stripped of their furcation tubing and buffer and are placed in the connector with a pitch of 125\,$\mum$.
After 30\,cm, the \ac{SMF} and the \acp{MMF} separate into two individual 5\,m-long fiber cables: 1)\,the science \ac{SMF}, which is terminated to an FC/PC adapter to feed the science instrument and 2)\,the sensing \acp{MMF}, which are rearranged into a linear array within an SMA connector.

The \ac{SMF} (Fibercore SM980) features \iac{MFD} of 5.8\,$\mum$ ($\mathrm{1/e^{2}}$-intensity at 980\,nm) and is taken from the same batch of the fiber that will feed the iLocater spectrograph, minimizing any fiber-to-fiber coupling losses further down the fiber link.
To simplify design and production, the \acp{MMF} are off-the-shelf fibers (Thorlabs FG105LCA). Their optical properties (core diameter 105\,$\mum$, NA=$0.22$) were chosen in order to reduce the core-to-core separation between the \ac{SMF} and \acp{MMF}, reducing the 3D printed lens dimensions.

\subsection{Lens design}
\label{sec:methods-design}

Design and optimization of the \ac{MLR} were performed using the optical design software \textit{Zemax OpticStudio}.
To calculate the coupling efficiency into the \ac{SMF}, the \ac{POP} tool was employed, and for \ac{MMF} coupling the \textit{Imaging} tool was used. \ac{POP} uses Fourier and Fresnel propagation, which is crucial when handling the near-Gaussian mode of the \ac{SMF} and the complex illumination pattern on the \ac{MLR}. It is computationally intensive however, so to design the shape of the lenses, the \textit{Imaging} tool was used, which utilizes a ray tracing algorithm to estimate the coupling efficiency into \acp{MMF}.

For our technology demonstrator, we aimed to have a strong signal for tip-tilt sensing while also enabling high \ac{SMF} coupling efficiency. This will both increase the \ac{SNR} and also provide a signal in all six fibers within a reasonable dynamic range.
The diameter of the central aperture was chosen to clip $\sim$13\% of the light, reducing the maximum achievable \ac{SMF} coupling efficiency with an idealized circular pupil from $~\sim$80\% \cite{Shaklan1988} to $\sim$65\%.
Using this aperture, the surface shape of the \ac{MLR} was then optimized to maximize the \ac{MMF} coupling efficiency, weighted to favor on-axis beams with decreasing priority for misalignment up to $100\,\mum$ (corresponding to $\sim$20\,$\lambdaD$).
The surface shape of the individual lenses needs to provide suitable optical power to focus the incoming clipped part of the beam into the \ac{MMF}. This was achieved by optimizing the spherical shape and then adding corrections with both Zernike focal sag and separate conical constants in both directions.
A strong optical power was necessary to refract the beam from the inner edge of the microlens to the \ac{MMF}.
For this, polynomial corrections were successively applied up to fourth order in the axis parallel to the radial axis, no additional correction was applied in the angular direction.

\subsection{Lens manufacturing}
\label{sec:methods-manufacturing}

The \ac{MLR} was manufactured using two-photon polymerization using a proprietary resin on the fiber tip \cite{Dietrich2018}, which allows the manufacturing of free-form lenses on small scales. Due to the use of stages in the printing process, these structures can take arbitrary shapes, limited by the need for an appropriate support structure and macroscopic forces.
The printing is aided by back-illuminating the fiber bundle and yields sub-micron alignment precision \cite{Dietrich2017} compensating for irregularities in the bundle geometry.
The process allows a precision of $\sim$100\,nm and \iac{RMS} surface roughness of $\sim$10\,nm. The physical size was limited to the maximum build height of approximately $400\,\mum$, due to the manufacturing stages and microscope objective \ac{NA}.

Once the \ac{MLR} was printed on the fiber the FC/PC connector was then placed within a bulkhead adapter (Thorlabs HAFC) for mechanical protection.

\subsection{Laboratory sensor response}
\label{sec:lab-response}

\begin{figure}[htb]
  \centering\includegraphics[width=\columnwidth]{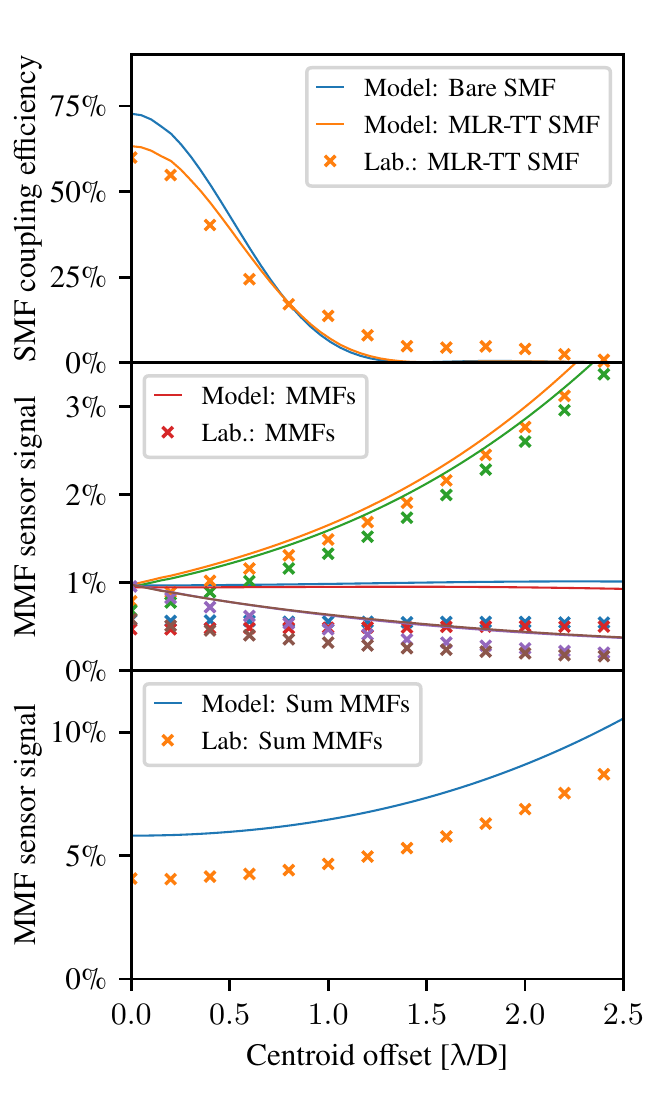}
  \caption{Modeled (solid lines) and measured (crosses) sensor response as function of centroid offset.
  \textbf{Top panel:} The coupling efficiency of the science SMF.
  \textbf{Middle panel:} The response of the six sensing MMF as function of beam offset.
  \textbf{Bottom panel:} MLR-TT sensor signal summed over all six \acp{MMF}.
  }
  \label{fig:sensor-response}
\end{figure}

As the custom lenses belonging to the iLocater acquisition camera were unavailable for laboratory experiments, the MLR-TT sensor's response was tested using commercial lenses. \Iac{SMF} illuminated by a 1050\,nm SLED source (Thorlabs S5FC1050P), was apertured and a Thorlabs AC127-025-C lens was used to produce an NA of 0.14, simulating the telescope's Airy disc. The experimental system provided a lower throughput than the final on-sky experiment, due to lower image quality. The results in Fig.~\ref{fig:sensor-response} show the sensor's response to an gradually off-centered beam in the laboratory setup, both as modeled and as measured.
The modeled \ac{SMF} coupling efficiency (Fig.~\ref{fig:sensor-response}, top) includes Fresnel reflection loss of 3.5\% at both fiber input and output face. The maximal achievable coupling efficiency within the MLR-TT sensor's \ac{SMF} is measured at $59.9\pm0.6\%$ which is slightly lower than the expected value of 63.2\% at the given wavelength. This coupling efficiency then drops off slightly faster than expected with an off-centered beam but features a slightly increased coupling for misalignment of up to $2.2\,\lambdaD$. The causes of this behavior still to be understood but are likely due to fiber bundle and lens imperfections.

The response of the sensing \acp{MMF} (Fig.~\ref{fig:sensor-response}, center) follows the modeled curves well, though the six sensing \acp{MMF} are not evenly illuminated when the beam is centered. During alignment we found that the illumination pattern depends strongly on the fiber alignment angle (pitch and yaw) and could not be completely corrected. This can result from asymmetries in the beam or uneven \ac{MMF} properties such as irregular spacing or different fiber losses. In practice this is corrected by the calibration routine (Sec.~\ref{sec:methods-recon}).

Laboratory results show the \ac{MLR} couples 4.1\% of the overall light into the \acp{MMF} when the beam is centered, which is 30\% lower than the modeled value of 5.8\% (this includes 11\% reflections and losses from the fiber and 8\% from the lens).
Interestingly, this loss remains constant with respect to beam position (Fig.~\ref{fig:sensor-response}, bottom) up to a centroid offset of $~\sim3\,\lambdaD$.
We presume that the remaining mismatch is due to a non-optimally shaped lens surface. The ray approximation as described in Sec.~\ref{sec:methods-design} only considers a central top-hat beam but fails to accurately account for the diffractive pattern that illuminates the lenses outside the central beam.

Theoretical throughput calculations and the corresponding photon, sky background and camera noise associated with the described system show that with this reduced sensor signal, a source with 8$^{\rm{th}}$\,magnitude in the J band can provide a \ac{SNR} of 14 for each \ac{MMF} output when running at 500\,Hz.
Simulations with the same pipeline as described in Sec.~\ref{sec:methods-simulations} show that this results in an reconstruction accuracy of $\sim$0.1\,$\lambdaD$ in tip and tilt combined.
In this limiting case, performance is limited by read-out noise of the detector.

\subsection{Signal processing}
\label{sec:methods-processing}
The output of the sensing \acp{MMF} was re-imaged with two lenses mounted within a hybrid tube and cage mechanical system and directly attached to the lens interface of a First Light C-Red 2 InGaAs detector.
This detector was chosen as it provides both a high frame rate (up to 16\,kHz) and low read-out noise (34\,e$^{-}$) with a pixel size of 15\,$\mum$.
Each \ac{MMF} illuminates a circular region on the detector with a diameter of 100\,$\mum$.
For each fiber, the 20 pixels with the highest \ac{SNR} are selected and used for further processing.
In laboratory tests, 20 pixels were measured to provide a steady fraction of $80\%$ of the flux and the best overall \ac{SNR}.
The detector data was then processed by the \ac{DARC} \cite{Basden2010,Basden2012}, running on a consumer grade desktop computer.

\subsection{Reconstruction and calibration}
\label{sec:methods-recon}

\begin{figure*}[htbp]
  \centering
  \includegraphics[width=\textwidth]{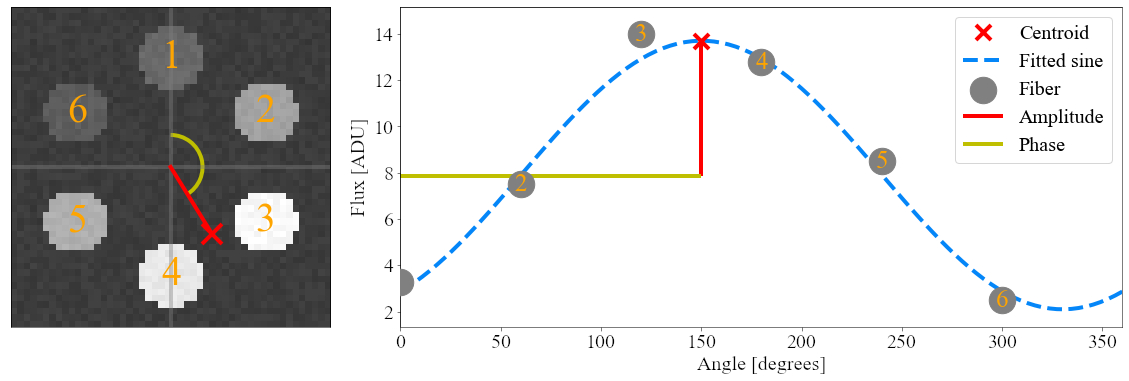}
  \caption{Illustration of the reconstruction routine with simulated noise. \textbf{Left panel:} Simulated detector image showing the six \acp{MMF} (numbered 1-6 in orange) along with the reconstructed centroid of the \ac{PSF} (red cross). \textbf{Right panel:} A graphical illustration of the reconstruction routine. Here the six fiber fluxes (gray, numbered 1-6 in orange) are ordered by their azimuthal coordinate and a sine function with angular period of $2\pi$ is fitted, giving the angle, amplitude and offset of the centroid.}
  \label{fig:reconstruction_simple}
\end{figure*}

The reconstruction algorithm (see Fig. \ref{fig:reconstruction_simple}) calculates the \ac{MMF} illumination and converts it to a physical centroid position.
For this, the six fiber fluxes are ordered with their azimuthal coordinate and a sine function with angular period of $2\pi$ is fitted to this signal.
Three best fit parameters are obtained by this routine (see Fig.~\ref{fig:reconstruction_simple}):
\begin{enumerate}[itemsep=0mm]
    \item Offset, depending on both background signal and target flux.
    \item Amplitude, corresponding to the radial position of the beam. Note, this is an arbitrary flux unit and the amplitude does therefore not directly yield the physical centroid position.
    \item Phase, corresponding to the azimuthal coordinate of the centroid position.
\end{enumerate}

Laboratory tests showed that this approach yields the most reliable and stable output, less susceptible to noise than a simple \ac{CoM} algorithm.

A calibration routine is used to correct the reconstructed centroid position for accurate loop feedback and run time diagnostics.
It accounts for irregularities in the system such as asymmetries or misalignment of the \ac{MLR}, transmission variations within the fiber bundle and static aberrations in the \ac{PSF}.
For this, a circular motion is introduced with the tip-tilt mirror.
The offset between the introduced and reconstructed azimuthal coordinate and the factor between the respective radial coordinates is approximated with individual best fit \acp{DFT} of $5^{\mathrm{th}}$ order as a function of the azimuthal coordinate.
The obtained correction function is subsequently applied to the measured centroid position.
It should be noted that this calibration routine is repeated for each target in order to remove slowly changing quasi-static aberrations (arising from effects such as mechanical flexure) and to include asymmetries of the source itself such as companions or background sources.

The interaction matrix is constructed by applying a linear signal in both tip and tilt with the mirror and simultaneously measuring the centroid position.
The resulting 2x2 matrix is then inverted to obtain a reconstruction matrix, which can be used by the control loop to convert the measured centroid position into an feedback signal to command the tip-tilt mirror.

\subsection{On-sky integration}
\label{sec:methods-integration}

\begin{figure}[htb]
  \centering
  \includegraphics[width=.8\columnwidth]{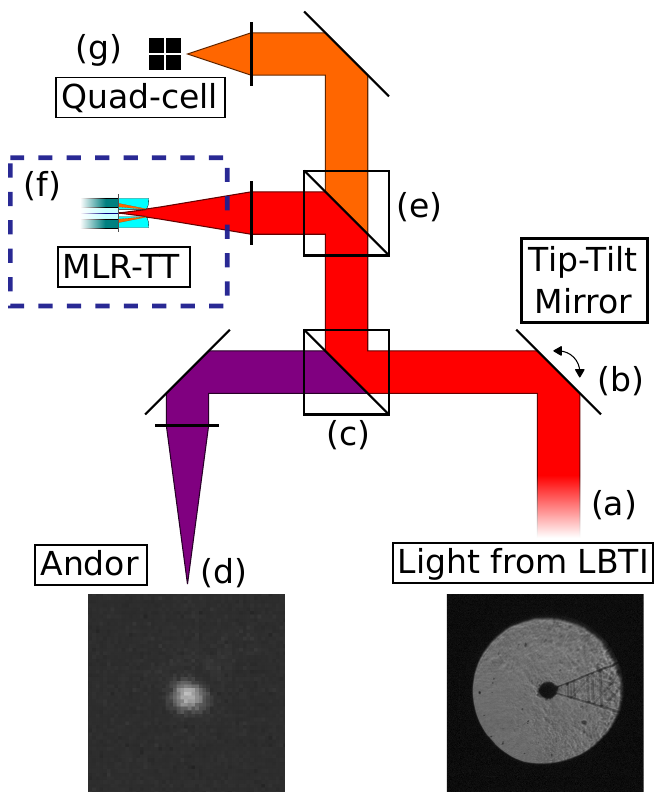}
  \caption{Optical path of the experimental setup with the iLocater acquisition camera at \ac{LBT} (sizes are not to scale).
  (a) The collimated \ac{AO} corrected beam from \ac{LBTI} is steered by a fast tip-tilt mirror (b).  A short-pass dichroic (c) transmits wavelengths between 920 and 950\,nm to  be imaged by the Andor focal plane camera (d).
  The science light is reflected by the long-pass dichroic mirror (e) and focused into  the MLR-TT sensor (f) and \ac{SMF}.
  Light between 1.34-1.76\,$\mum$ is transmitted and imaged on  the quad-cell (g) that can be used in closed-loop to correct for tip-tilt vibrations.}
  \label{fig:optical-path}
\end{figure}

The MLR-TT sensor was integrated into the iLocater SX acquisition camera \cite{Crass2020} that is fed by the \ac{LBTI}.
The optical path is illustrated in Fig.~\ref{fig:optical-path}.
The iLocater acquisition camera receives the pupil from the telescope (a), passes the wavelengths between 920\,nm and 950\,nm (c) to its imaging channel equipped with an Andor focal plane camera (ANDOR Zyla 4.2 Plus, d), providing a sampling of 6.1 pixels across the \ac{FWHM} of the diffraction-limited \ac{PSF}. This focal plane image is used as reference for the centroid position, i.e. the tip-tilt.

iLocater's native tip-tilt correction features a quad-cell photo detector (Hamamatsu G6849-01 InGaAs, g), which is fed with light picked off by a dichroic at 1.34-1.76\,$\mum$, (e) just before the final coupling optics. The quad-cell system can then feed an error signal back to a fast tip-tilt mirror (nPoint RXY3-276, b) to correct for tip-tilt.
Alternatively, the mirror can be controlled by the MLR-TT sensor to either introduce the required motions for calibration (see Sec.~\ref{sec:methods-recon}) or for correcting tip-tilt directly.

The science beam ($0.97-1.31\,\mum$) is focused by two custom triplet lenses \cite{Crass2020} to an f/3.7 beam on the \ac{SMF} to match its \ac{MFD} of 5.8\,$\mum$ ($\mathrm{1/e^{2}}$-intensity at 970\,nm). The fiber mount can be moved in 5 axes for alignment and to switch between three independent fibers mounted at the instrument focal plane. These are: the native iLocater \ac{SMF}, a bare MMF ($105\,\mum$ core diameter) used for flux calibration, and the guest fiber port equipped with the MLR-TT sensor (f).

Fiber throughput is determined by measuring the output flux from each fiber with the bare \ac{MMF} serving as an incident flux reference. Output flux is measured with a FemtoWatt receiver \cite{Crass2020}.
The fiber bundle holding the six sensing \acp{MMF} is routed to a separate opto-electric enclosure, housing the read-out optics and electronics.

\subsection{Simulations of on-sky results}
\label{sec:methods-simulations}

To further investigate the performance of the sensor with our recorded on-sky conditions, we simulated the sensor response for differing \ac{AO} correction. To do this, an atmospheric wavefront distortion of 1000 modes in combination with a corresponding \ac{AO} system correcting 500 modes was modeled using the HCIPy high contrast imaging simulation framework \cite{por2018hcipy}. To allow an accurate comparison, the tip and tilt modes of the resulting wavefront are replaced by the centroid positions that were recorded during the on-sky observations.

These simulations are key as they allow us to understand our results and estimate the impact of residual \ac{AO} aberrations and their dominance with respect to other noise sources.

\section{Results}
\label{sec:results}

We tested the MLR-TT on-sky in November 2019 at the \ac{LBT}, using the left (SX) mirror of the telescope \cite{Crass2020}. During the run the \ac{LBTI-AO} system was using the SOUL upgrade, which is designed to produce \iac{SR} of up to 78\% in I-band \cite{Pinna2016} under optimal conditions. For all observations the \ac{AO} system was running at 1 kHz closed on 500 modes. Correction for \ac{AO} \ac{NCPA} was performed before observations, but otherwise there was no direct interaction between the MLR-TT sensor and \ac{LBTI-AO}.

We present the results from three on-sky targets, with a total of 8 datasets. All targets were chosen to be bright (< 6$^{\rm{th}}$ magnitude), marginalizing  detector noise from the MLR-TT sensor. Tab.~\ref{tab:targets} provides an overview of the targets, the \ac{AO} loop performance, and the associated datasets.

\begin{table}[htb]
\centering
\caption{\bf Observed targets and datasets as well as observational seeing, estimated \ac{SR} and and the status of the tip-tilt correction loop.}
\begin{tabular}{|p{1.3cm}p{0.4cm}|p{0.58cm}|p{1.0cm}|p{1.3cm}|p{1.4cm}|}
\hline
{\bf Target}/dataset & &
        J-band mag.&
        Seeing ($''$) &
        Est. SR &
        Additional Tip-Tilt control
        \\
\hline
  \textbf{HIP28634}
  & /4 & 5.3 & 1.2-2.0 & $50\pm6\%$ & MLR-TT \\
  & /5 & $''$ & $''$ & $52\pm7\%$ & None \\
  \hline
  \textbf{HD12354}
  & /1 & 5.9  & 1.0-1.4 &  $67\pm7\%$  & None\\
  & /2 & $''$ & $''$    &  $67\pm11\%$ & MLR-TT \\
  \hline
  \textbf{HIP7981}
  & /2 & 3.8 & 1.0-1.4 & $66\pm4\%$ & MLR-TT \\
  & /4 & $''$ & $''$  & $65\pm4\%$ & MLR-TT \\
  & /5 & $''$ & $''$  & $65\pm4\%$ & MLR-TT \\
  & /6 & $''$ & $''$  & $65\pm4\%$ & MLR-TT \\
\hline
\end{tabular}
  \label{tab:targets}
\end{table}

Each dataset includes three simultaneous measurements taken using iLocater and the MLR-TT sensor:
\begin{itemize}
    \item Andor focal plane frames (Sec.~2.\ref{sec:methods-integration}), taken at a frame rate of 250 Hz. A symmetric 2D Gaussian function is fitted to the data in post processing and its calculated centroid used as a reference for \ac{PSF} position. The \ac{SR} in Tab. \ref{tab:targets} was estimated by fitting a Gaussian to the centroid corrected \ac{PSF} and taking the ratio between the normalized central intensities of this fit and the expected telescope \ac{PSF} as described in Ref.~\citenum{Bechter2019}. Due to the limited \ac{SNR} of the individual frames, the \ac{SR} calculations were smoothed by applying a moving median algorithm covering 20 frames.
    \item The reconstructed centroid position from the MLR-TT sensor (Sec.~2.\ref{sec:methods-recon}). Data were taken at a frame rate of 500\,Hz. In post processing the frames were interpolated and cross-correlated to match the time reference of the Andor data.
    \item The \ac{SMF} coupling efficiency was measured with the FemtoWatt receiver (Sec.~2.\ref{sec:methods-integration}).
\end{itemize}

\subsection{Sensor calibration}
\label{sec:results_calibration}

\begin{figure}[htbp]
  \centering
  \includegraphics[width=\columnwidth]{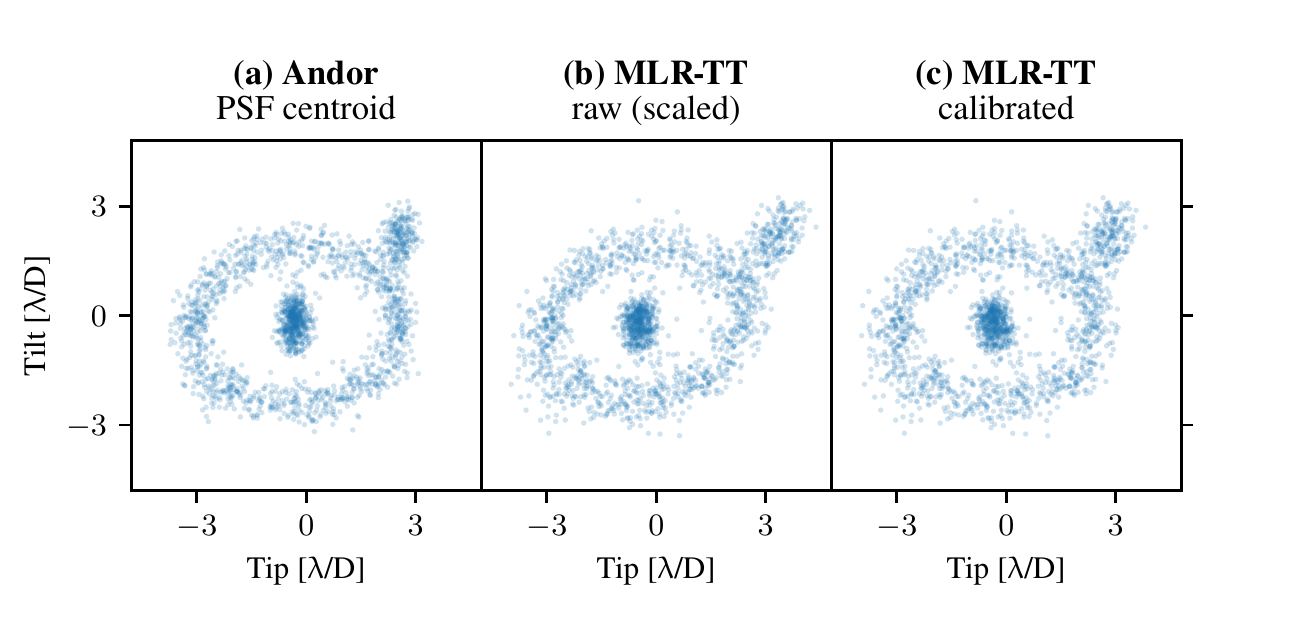}
  \caption{Three scatter plots showing the on-sky calibration routine of target HIP7981. Shown are (a) the reference centroid position measured with the Andor focal plane camera, (b) the MLR-TT reconstructed raw centroid position reconstructed from the MLR-TT  and (c) the calibrated MLR-TT centroid position (see Tab.~\ref{tab:calibration-rms}).
  Initially the \ac{PSF} is centered, then a circular motion is introduced on the fast tip-tilt mirror. This movement is not calibrated in $\lambdaD$ and produces an elliptical shape in the focal plane due to the angle of the tip-tilt mirror. The introduced figure also features a central accumulation from before and after the circular motion, as well as an introduced step position seen as a separate patch to the top right of the circle.}
  \label{fig:recon_calibration}
\end{figure}

As described in Sec.~2.\ref{sec:methods-recon}, the calibration pattern was generated by introducing a circular motion on the tip-tilt mirror by issuing open loop position commands. An example of the calibration routine for target HIP7981 is shown in Fig.~\ref{fig:recon_calibration} for (a)\,the Andor reference centroid position, (b)\,the raw MLR-TT centroid position and (c)\,the calibrated centroid position.

During the calibration, the \ac{AO} loop was closed, but no additional tip-tilt correction was applied. Due to residual vibrations at the telescope, the measured centroid positions show a broadened pattern, which is  averaged. The averaged centroid positions are used to correct the reconstructed centroid for static asymmetries.

For HIP7981, the reconstruction without calibration shows an \ac{RMS} error of  0.33\,$\lambdaD$ in tip and 0.26\,$\lambdaD$ in tilt (0.42\,$\lambdaD$ combined).
After correction, this improves to 0.19\,$\lambdaD$ in tip and 0.21\,$\lambdaD$ in tilt (0.28\,$\lambdaD$ combined) and appears random.
The impact of the calibration on the reconstruction accuracy for all targets is listed in Tab.~\ref{tab:calibration-rms}, including the \ac{RMS} shift that is applied by the calibration.
This shift corresponds to the correction that the calibration routine performs on the centroid position which is seen as an improvement of the reconstructed centroid position.
The correction is seen to provide a more significant improvement for the datasets with lower pre-calibration \ac{RMS} reconstruction error.
This arises from a more precise measurement of the calibration pattern (corresponding to a thinner ring in Fig.~\ref{fig:recon_calibration}) that leads to a more accurate parametrization of the correction function.

For all other datasets listed in Tab.~\ref{tab:targets}, the calibration was also applied but did not provide a significant improvement.
These datasets all feature a smaller dynamical range and the applied shift varied between $0.06$ and $0.09\,\lambdaD$ in tip and tilt combined.
Compared to the overall noise in these datasets (see Sec.~\ref{sec:results_ao}), the impact of the calibration is negligible.

\begin{table}[htb]
\centering
\caption{\bf Improvement gained through the calibration routine. \ac{RMS} reconstruction error before and after applying the calibration is listed as well as the RMS shift determined after the application of the calibration routine.}
\begin{tabular}{|p{1.0cm}p{0.5cm}|p{1.4cm}|p{1.4cm}|p{1.4cm}|}
\hline
{\bf Target} & &
        RMS error no calib. [$\lambdaD$] &
        RMS error calibrated [$\lambdaD$] &
        RMS calib. shift [$\lambdaD$]\\
\hline
  \textbf{HIP28634}
  &/cal.  & 0.54 & 0.50 & 0.23 \\
  \hline
  \textbf{HD12354}
  &/cal. & 0.42 & 0.31 & 0.27 \\
  \hline
  \textbf{HIP7981}
  &/cal. & 0.42 & 0.28 & 0.30 \\
\hline
\end{tabular}
  \label{tab:calibration-rms}
\end{table}

\subsection{Closed-loop performance}
\label{sec:results_loop}
\begin{figure}[htb]
  \centering
  \includegraphics[width=\columnwidth]{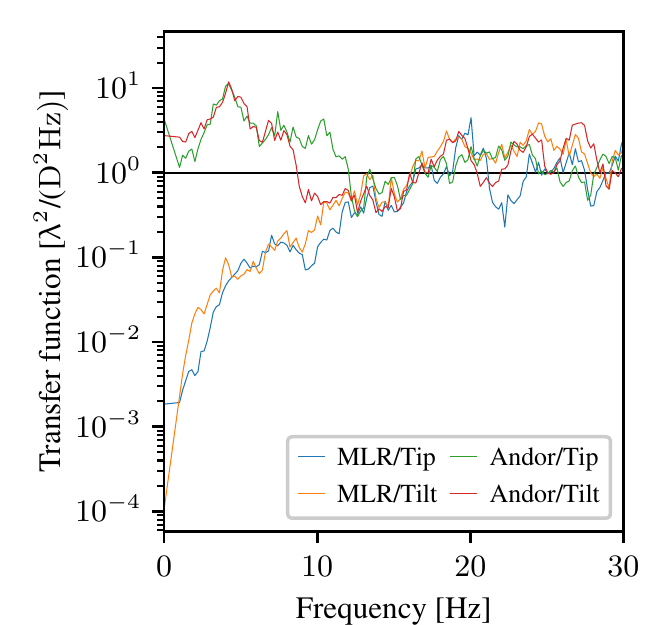}
  \caption{Closed loop transfer function when stabilizing the beam with the MLR-TT sensor for
  datasets HIP28634/4 (closed loop) and HIP28634/5.
  Below $\sim$10\,Hz, the MLR-TT sensor (blue, yellow) detects a different frequency rejection than the Andor reference (green, red), whilst above 10\,Hz the transfer functions agree well. }
  \label{fig:loop_rejection}
\end{figure}

In the datasets listed in Tab.~\ref{tab:targets}, the acquired \ac{PSF} centroid positions were used to drive the tip-tilt mirror.
While the loop was operating stably, no improvement in \ac{SMF} coupling was observed.
The closed loop transfer function as seen by the MLR-TT sensor (Fig.~\ref{fig:loop_rejection}, blue/orange) shows a significant rejection of frequencies below 15\,Hz, however this is not seen in the Andor reference camera (Fig.~\ref{fig:loop_rejection}, green/red).
Above 15\,Hz both Andor and MLR-TT show the same behavior, however the loop fails to correct for the faster disturbances. This suggests that the loop is not running at a high enough frequency for correction or the latency is too high.

\subsection{Reconstruction accuracy}
\label{sec:results_reconstruction}

\begin{figure}[htb]
  \centering\includegraphics[width=\columnwidth]{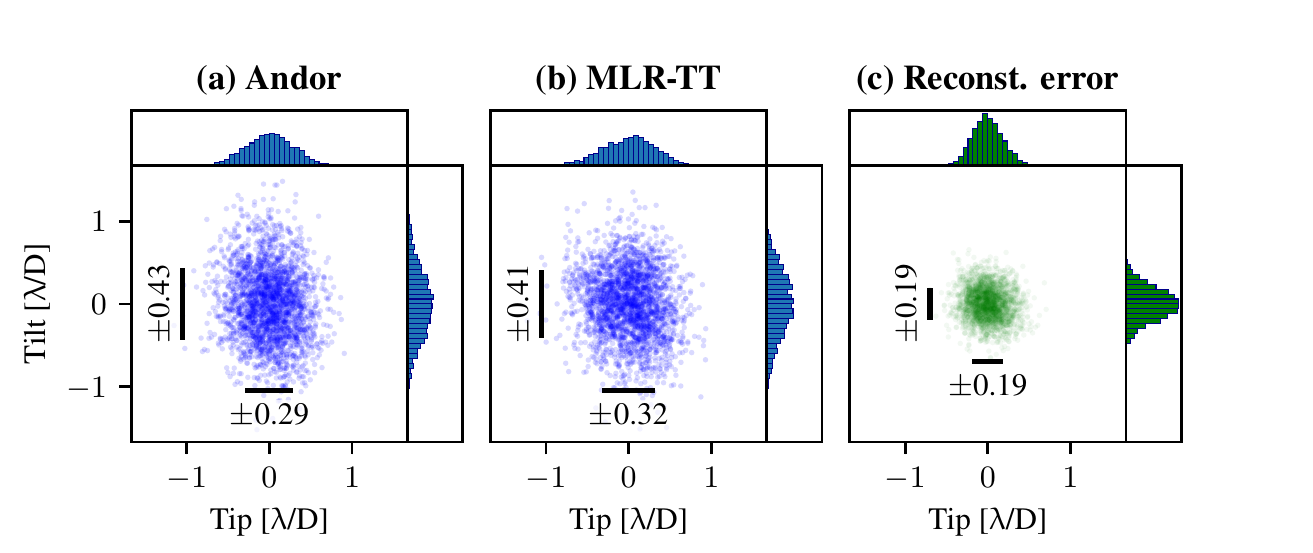}
  \centering\includegraphics[width=\columnwidth]{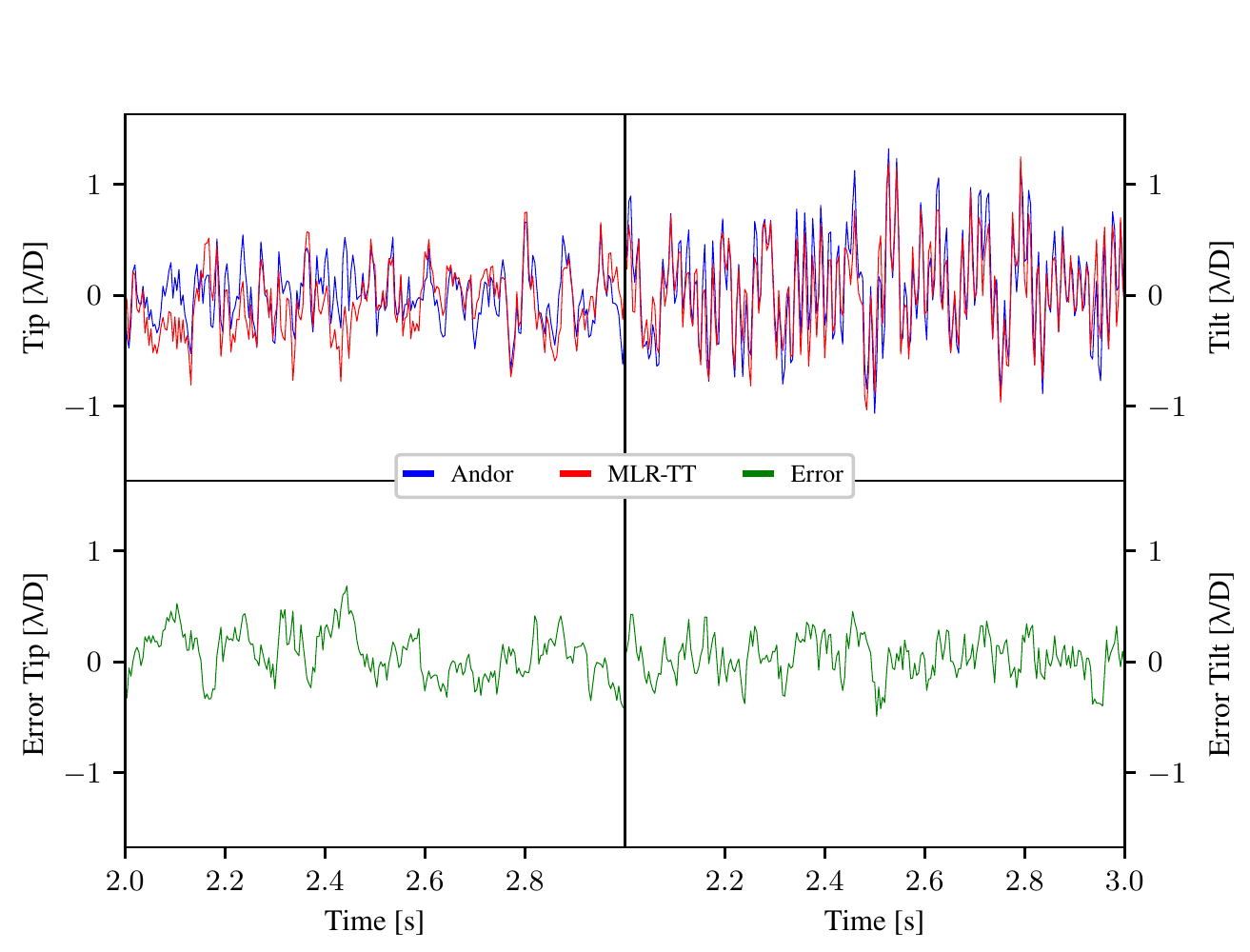}
  \caption{Sensor reconstruction accuracy shown graphically. \textbf{Top panels:} Time series of (Left) the centroid position measured by the Andor focal plane camera, (Center) the position reconstructed with the MLR-TT and (Right) the error in the reconstruction by taking the difference between the former two datasets. \textbf{Bottom panels:} Time series graphs of the same dataset for: (Top) Comparison of the centroid $x$-position for (tip, left) and $y$-position (tilt, right) of Andor reference (blue) and MLR-TT (red), and (Bottom) the corresponding reconstruction error (green) from their difference.}
  \label{fig:recon_scatter}
\end{figure}

This significant mismatch between MLR-TT sensor and Andor reference in evaluating the loop performance needs to be understood.
For this, we analyze the accuracy with which the sensor is able to reconstruct the centroid position.
Fig.~\ref{fig:recon_scatter} shows the centroid position for the Andor reference and the MLR-TT sensor for HD12354/1, as well as the corresponding reconstruction error. While the scatter of these values does not show any systematic patterns, the time series (cutout, bottom) shows that the sensor is indeed able to track the centroid position.
The residual error features a mismatch, amounting to $0.19\,\lambdaD$ \ac{RMS} in both tip and tilt.

\begin{figure}[htbp]
  \centering
  \includegraphics[width=\columnwidth]{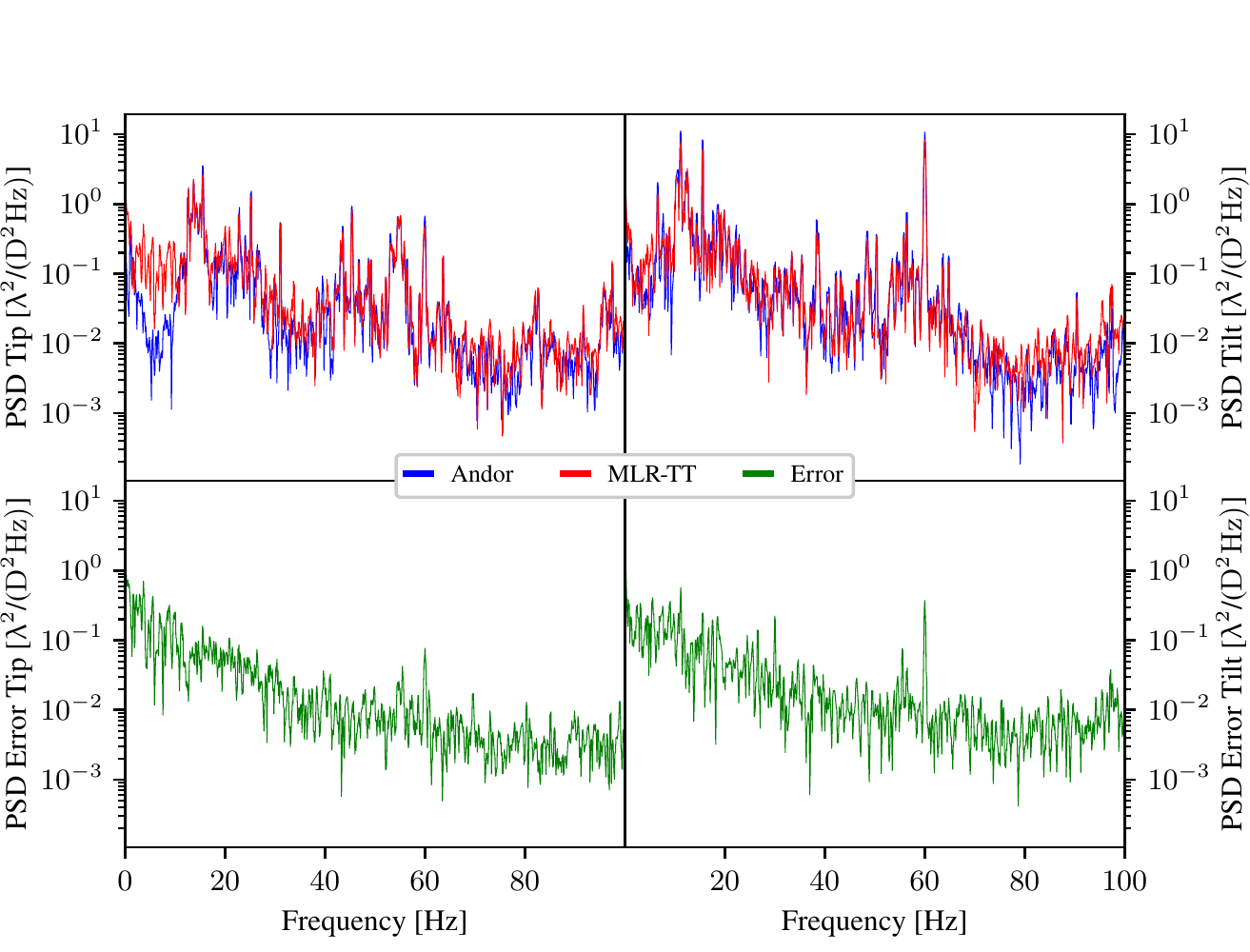}
  \caption{Power spectral density (PSD) for HD12354/1 of the signal shown in Fig.~\ref{fig:recon_scatter}. (Top) \ac{PSD} of centroid $x$-position (tip, left) and $y$-position (tilt, right) of MLR-TT (red) compared to the Andor reference (blue), (Bottom) the \ac{PSD} of the corresponding reconstruction error (green) for tip and tilt. Most of the vibrational power lies between 10 and 20\,Hz, whilst most of the reconstruction error is in the low frequencies ($<10$\,Hz).}
  \label{fig:recon_psd}
\end{figure}

The time series of the error suggests a strong low-frequency component.
The \ac{PSD} of the MLR-TT sensor tracks this behavior very well (see Fig.~\ref{fig:recon_psd}), with the sensor \ac{PSD} tracking the features of the reference centroid very accurately above 10\,Hz.
Residuals below 10 Hz are calculated to account for approximately 50\% of the combined tip-tilt error, while residuals between 10 and 20 Hz contribute less than 20\%.

\subsection{Impact of AO performance}
\label{sec:results_ao}

\begin{figure}[htb]
  \centering\includegraphics[width=\columnwidth]{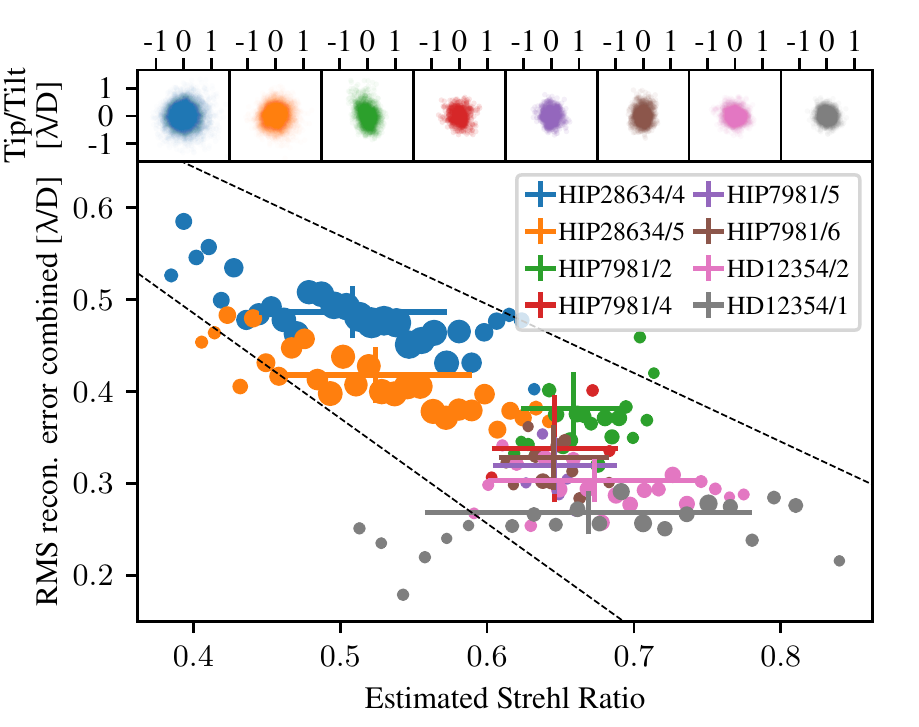}
  \caption{On-sky sensor performance. \textbf{Main panel:} Reconstruction accuracy as a function of estimated \acf{SR} for these datasets. Cross marks in the main plot represent the mean and error for each dataset, while the circles in the subplots correspond to subsets with different estimated \acp{SR}, with the size of the circle representing the number of frames in each set. The dashed lines show the fitting error. \textbf{Top panels:} The centroid reconstruction error scatter plot for each analyzed dataset.}
  \label{fig:recon_srrms}
\end{figure}

Fig.~\ref{fig:recon_srrms} shows the combined tip-tilt reconstruction error for all datasets as a function of estimated \ac{SR}. Note that all datasets feature similar \ac{RMS} centroid values ($\sim$$\lambdaD$).

The wavefront correction varies significantly between the  datasets and within individual datasets, with subsets featuring \acp{SR} as low as 40\% and reaching up to 80\%.
The reconstruction accuracy shows a strong dependency on the \ac{SR} and improves significantly with increasing \ac{SR}.
The best reconstruction shows a combined tip-tilt \ac{RMS} of  0.27\,$\lambdaD$ while the worst reconstruction reaches \iac{RMS} error of 0.5\,$\lambdaD$.
A linear fit yields a slope of $-0.95\pm0.20\,\lambdaD$, an improvement in \ac{RMS} reconstruction accuracy of $\sim$0.1\,$\lambdaD$ per $10\%$ increase in \ac{SR}.

\subsection{On-sky sensor simulations}
\label{sec:results_aosim}

\begin{figure}[htb]
  \centering\includegraphics[width=\columnwidth]{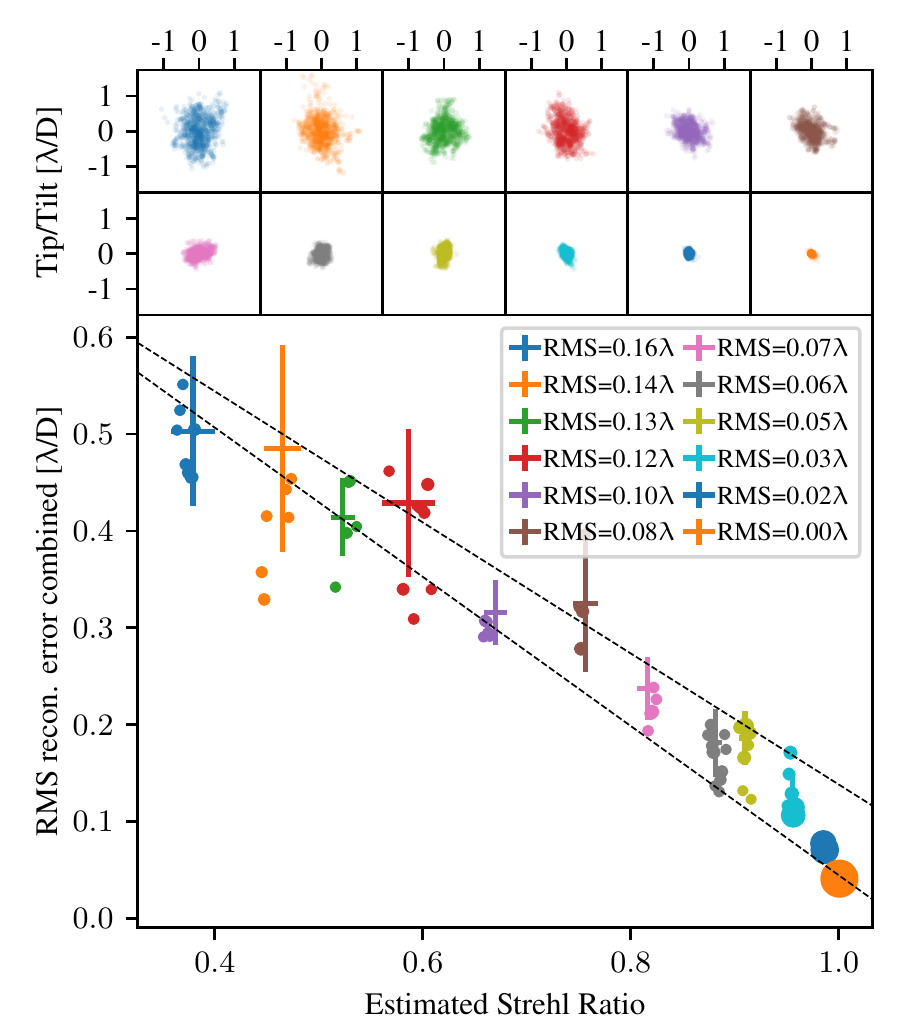}
  \caption{Synthetic MLR-TT sensor performance derived from AO simulations, plotted to be comparable to Fig.~\ref{fig:recon_srrms}.
  \textbf{Main panel:} Reconstruction accuracy as a function of \acf{SR} for \ac{AO} simulations with varying residual aberration strength labeled with their \ac{RMS} wavefront error. Crosses represent overall mean and error for each data set, while the circles correspond to subsets binned by \ac{SR}, with the size of the circle representing the number of frames in each set.The dashed lines show the fitting error. \textbf{Top panels:} Centroid reconstruction error scatter for the individual datasets.}
  \label{fig:aosim_srrms}
\end{figure}

\ac{AO} simulations as described in Sec.~\ref{sec:methods-simulations} were performed to reconstruct the sensor operation. Fig.~\ref{fig:aosim_srrms} shows the resulting reconstruction error for tip and tilt combined as a function of the retrieved \ac{SR} and is analogous to Fig.~\ref{fig:recon_srrms}.
For the lowest simulated \acp{SR} of $\sim$$50\%$, reconstruction accuracy is worse than $0.35\,\lambdaD$ and improves to $0.16\,\lambdaD$ for \iac{SR} of 80\%.
As with the on-sky results (cf. Fig.~\ref{fig:recon_srrms}), the data are well fit by a linear trend, with a slope of $-0.72\pm0.05\,\lambdaD$. For completeness, we have also simulated the reconstruction error for a flat wavefront (Fig.~\ref{fig:aosim_srrms}, yellow marker) which shows a reconstruction error of less than $0.05\,\lambdaD$.

\section{Discussion}
\label{sec:discussion}

In the preceding section we presented the on-sky performance of the MLR-TT sensor. Whilst able to track incident beam motions, the sensor was unable to improve fiber coupling performance with our current \ac{AO} loop. The sensor also shows limitations in the overall performance which can be achieved due to the effects of residual aberrations. The causes and solutions are discussed in this section.

\subsection{Sensor reconstruction limitations}
\label{sec:discussion-performance}

As shown in Fig.~\ref{fig:recon_srrms}, the sensor was able to reconstruct the centroid position to an accuracy of 0.27 \,$\lambdaD$ combined tip-tilt \ac{RMS}. The majority of this error (50\%) originates in frequencies below 10\,Hz and depends strongly on the estimated \ac{SR}. To ascertain the cause of this error, we presented optical simulations with differing \ac{SR} in Sec.~3.\ref{sec:results_aosim}.
The simulations show the same trend with a slightly flatter linear fit.
The discrepancy can by attributed to a number of additional noise sources that occur within the measurements.
These alternative sources include detector noise, reconstruction algorithm error, \ac{NCP} vibrations, flux variations, and noise in the measurements of the reference centroid.
While we investigated these factors during analysis, the current system is most strongly impacted by the effects of residual aberrations.
For future versions of the sensor we aim to understand the exact contributions that these noise terms have on the reconstruction accuracy.

To further investigate the impact of wavefront aberrations on the MLR-TT sensor, in future laboratory testing and on-sky experiments, we intend to acquire additional metrology data to identify other effects driving performance. This will allow us to optimize the MLR-TT reconstruction algorithm to account for the observed aberrations and possibly even reconstruct Zernike modes beyond tip and tilt.

\subsection{Loop performance}
As illustrated in Fig.~\ref{fig:recon_srrms}, under the best conditions experienced, the reconstruction accuracy of the sensor provided a combined \ac{RMS} error of $0.27\,\lambdaD$. Assuming an ideal control system, this would provide correction with an \ac{RMS} error 1.5 times lower than the existing quad-cell system. With our current control system, this is reduced significantly due to latency and meant the loop was only able to reject frequencies up to 15-20\,Hz.
The control system therefore needs to be optimized in order to allow a better correction of the tip-tilt disturbance which holds the most power in frequencies between 10 and 20\,Hz (see Fig.~\ref{fig:recon_psd}).

As shown in Fig.~\ref{fig:recon_psd}, most of the noise in the reconstruction occurs below 10\,Hz.
The main goal will be to optimize the MLR-TT sensor software (Sec.~\ref{sec:discussion-performance}) and hardware design (Sec.~\ref{sec:discussion-optimization}) to improve its performance in this regime.
Even without additional precision, the loop can be tuned to filter this frequency range or another sensor designed to supress vibrations in the range 1-10\,Hz can be added.
Alternatively, the MLR-TT sensor may be used to only detect slow beam drift below 1\,Hz. Any residual aberrations will average out over long timescales (>1 second) and the sensor can be optimized to measure slow mechanical drift resulting from e.g. gravitational flexures.
This would focus the sensor on utilizing one of its main advantage, namely that it is virtually free from \ac{NCP} effects. When running at lower frame rates, the sensor also needs less light for operation, increasing the limiting sensing magnitude and the light available for the science instrument.


\subsection{Sensor optimization}
\label{sec:discussion-optimization}

To control the amount of noise that is induced by residual \ac{AO} aberrations, the lens design can be tuned for future devices.
As the shape of the \ac{MLR} surface is set by the need to efficiently couple light into the \acp{MMF}, the height of the lens and the size of the central aperture then become the most important variables.
Both parameters control the distance from the focal plane where the telescope beam is sensed and by varying them the impact of aberrations in the system changes.

By sampling the beam closer to the fiber focal plane, the MLR-TT sensor will use an intensity distribution more similar to the \ac{PSF} for sensing, which depends mostly on the phase of the wavefront at the pupil. As the height of the \ac{MLR} increases, the beam enters the Fresnel regime  and the sensor is therefore also affected by variations in the pupil intensity that arise from scintillation and pupil instability. Fully analyzing this parameter space will be crucial for future sensor optimization.

The size of the lens ring aperture determines how much of the beam's central core is diverted to the sensor. As the edges of the beam are more susceptible to higher order modes and asymmetries, using more of the beam's core will result in more reliable measurements. However, this will also reduce the fraction of light available for science measurements.
This trade off is the key design choice that will be determined by future use cases and implementations.
In addition to the size of the central aperture, the \ac{NA} can be used to slightly change the ratio between sensor signal and \ac{SMF} coupling. Given the right optical system, it would be possible to perform individual adjustments of this trade off for each observed target.

\subsection {Future applications}

The system presented in this work was optimized to be used with the iLocater acquisition camera at the \ac{LBT}, however there are other diffraction-limited systems where the technology can find application.
As discussed in Sec.~\ref{sec:discussion-performance}, the performance is limited by residual \ac{AO} aberrations, and thus the most beneficial application will be with systems that feature as little residual wavefront aberrations as possible.

Besides current and future \ac{ExAO} system at large observatories, the MLR-TT sensor can have an advantage for small observatories, free-space optical communication systems and space based applications that employ diffraction-limited telescopes.
In these systems, the sensor can be integrated in a very compact fashion without the need for additional optical components in the optical train reducing complexity and mechanical footprint.

\section{Conclusion}
\label{sec:conclusion}
\acresetall

We presented the first on-sky results of our novel 3D-printed, fiber-based tip-tilt sensor (MLR-TT). The sensor was tested with the iLocater acquisition camera at the \acl{LBT} in November 2019.
The system consists of a 3D-printed \acl{MLR} that uses six \aclp{MMF} to reconstruct the centroid position, while providing an almost unobscured aperture where a science \acl{SMF} is positioned. This concept features a very small opto-mechanical footprint and degrades the maximum \acl{SMF} coupling efficiency by 15\%, which is comparable to typical losses due to beam aberrations.

We showed that the fundamental principle works well and  the sensor is able to reach a maximum reconstruction accuracy of $0.19\,\lambdaD$ in each tip and tilt, however, the system was not able to improve \acl{SMF} coupling efficiency.
The majority of the vibration was measured in frequencies between 10-20\,Hz, but the majority of the reconstruction error was shown to occur in low frequencies between 1-10\,Hz.
This error in reconstructing the centroid depended strongly on estimated \ac{SR} and subsequent simulations were able to recreate this trend, suggesting that residual aberrations were the dominating noise source that limited performance.

These findings will help to tune both the optical design and reconstruction algorithm to improve the centroid measurements and to reduce the impact of residual aberrations. Alternatively, the respective frequency range can be filtered or corrected using another sensor to minimize its impact.

We conclude that the MLR-TT sensor is best suited for applications requiring fast correction with low higher-order wavefront distortions while benefiting from its compact nature. This includes \acl{ExAO} systems, compact systems at small diffraction-limited telescopes and space based applications. We also note that the MLR-TT sensor operates very close to the fiber coupling surface, it is free of \acl{NCPA} and can therefore be used to track drifts and perform guiding in a closed-loop system where calibration between the wavefront sensor and fiber is difficult.

\section{Backmatter}

\begin{backmatter}

\bmsection{Acknowledgments}
P.H., R. J. H., and A.Q. are supported by the Deutsche Forschungsgemeinschaft (DFG) through project 326946494, 'Novel Astronomical Instrumentation through photonic Reformatting'.
This project has received support from the European Union's Horizon 2020 research and innovation program under grant agreement No 730890.
NAB, AGB and TJM acknowledge funding from UKRI Science and Technology Facilities Council (ST/T000244/1: NAB, TJM ; ST/P000541/1: AGB, NAB, TJM). \\
This research made use of HCIPy, an open-source object-oriented framework written in Python for performing end-to-end simulations of high-contrast imaging instruments \cite{por2018hcipy}, Astropy, a
community-developed core \texttt{Python} package for Astronomy \cite{astropy:2013, astropy:2018},
Numpy \cite{2020NumPy} and Matplotlib \cite{matplotlib}.

We thank the reviewers for the time improving this manuscript and  Romain Laugier for useful information about \ac{SMF} use in interferometry.

\bmsection{Disclosures} The authors declare no conflicts
of interest.

\bmsection{Data Availability Statement} Data underlying
the results presented in this paper are not publicly
available at this time but may be obtained from the
authors upon reasonable request.


\end{backmatter}

\bibliography{mlr-tt-library}
\end{document}